\documentclass[%
reprint,
groupedaddress,
amsmath,
amssymb,
aps,
prb,
floatfix,
]{revtex4-1}

\usepackage[T1]{fontenc}
\usepackage[usenames,dvipsnames]{color}
\usepackage{amsmath}
\usepackage{amsfonts}
\usepackage[caption=false]{subfig}
\usepackage{microtype}
\usepackage{graphicx}
\usepackage{grffile}

\newcommand{\JJ}{$J_1$-$J_2$-model}

\DeclareGraphicsExtensions{.pdf}

\begin{document}

\title{Incommensurability Effects in Odd Length $J_1$-$J_2$ Quantum Spin Chains: \\ On-site magnetization and Entanglement}

\author{Andreas Deschner} 
\email{deschna@physics.mcmaster.ca}
\author{Erik S. S{\o}rensen}
\affiliation{Department of Physics and Astronomy, McMaster University
1280 Main Street West, Hamilton, Ontario L8S 4M1, Canada}
\date{\today}

\begin{abstract} 
  For the antiferromagnetic $J_1$-$J_2$ quantum spin chain with an
  \emph{even} number of sites, the point $J_2^d=J_1/2$ is a disorder
  point. It marks the onset of incommensurate real space correlations
  for $J_2>J_2^d$.  At a distinct larger value of $J_2^L=0.52036(6)J_1$,
  the Lifshitz point, the peak in the static structure factor begins to
  move away from $k=\pi$. Here, we focus on chains with an \emph{odd}
  number of sites. In this case the disorder point is also at
  $J_2^d=J_1/2$ but the behavior close to the Lifshitz point,
  $J_2^L\simeq 0.538 J_1$, is quite different:   starting at $J_2^L$,
  the ground state goes through a sequence of level crossings as its
  momentum changes away from $k=\pi/2$.  An even length chain, on the
  other hand, is gapped for any $J_2>0.24J_1$ and has the ground state
  momentum $k=0$.  This gradual change in the ground state wave function
  for chains with an odd number of sites is reflected in a dramatic
  manner directly in the ground state on-site magnetization as well as
  in the bi-partite von Neumann entanglement entropy. Our results are
  based on DMRG calculations and variational calculations performed in a
  restricted Hilbert space defined in the valence bond picture. In the
  vicinity of the point $J_2=J_1/2$, we expect the variational results
  to be very precise.
\end{abstract}

\maketitle

\section{Introduction} \label{sec:intro} 
Disorder points were first discussed by Stephenson in models described
by classical statistical
mechanics.~\cite{Stephenson_1969,Stephenson_1970a,Stephenson_1970b,Stephenson_1970c}
On one site of a disorder point, the correlation function shows
monotonic decay, on the other oscillatory decay.  Depending on the how
the wavelength of the oscillation depends on the temperature, one
distinguishes between two kinds of disorder points. If the wavelength of
the oscillation depends on the temperature, one speaks of a disorder
point of the first kind, if it does not, one speaks of a disorder point
of the second kind.~\cite{Stephenson_1970a}
In the first studies, disorder points were found where the paramagnetic
phase of frustrated two-dimensional Ising models starts to show incommensurate
instead of commensurate behavior.
In models with competing commensurate and incommensurate order
one might expect such a point to occur where the short-range
correlations with the largest correlation length change from being
commensurate to being incommensurate. Such a point should then be
associated with a cusp in the correlation length, a fact that was
quickly established.~\cite{Stephenson_1970d} Schollw{\"o}ck, Jolicoeur
and  Garel first investigated disorder points in a quantum spin chain for
the bilinear-biquadratic $S=1$ quantum spin
chain,~\cite{schollwoeck_onset_1996} which has $H=\cos{\theta}\sum
\mathbf{S_i}\cdot\mathbf{S_{i+1}}+\sin{\theta}\sum\mathbf{S_i}\cdot\mathbf{S_{i+1}}$.
They pointed out that the disorder point in this gapped quantum model
coincides with the AKLT-point where $\tan{\theta_{VBS}}=1/3$ and the
known ground state is a valence bond solid (VBS) with correlation length
$\xi= 1/\ln(3)$.  They also identified another distinct point, the
Lifshitz point, at $\tan{\theta_L}=1/2$, where the peak in the structure
factor is displaced away from $\pi$ due to incommensurability effects. A
third distinct point in this $S=1$ model, $\tan{\theta_{disp}}\simeq
0.4$, has also been located~\cite{Golinelli_1999} where the minimum in
the magnon dispersion shifts away from $\pi$ and the curvature
(velocity) vanishes.  These 3 points can be distinct since no phase
transition occurs and the correlation length remains finite.
Subsequently it was confirmed~\cite{bursill_numerical_1995} that for the
$S=1/2$ $J_1$-$J_2$ spin chain with Hamiltonian: 
\begin{equation} 
  H=J_1\sum_i \mathbf{S}_i\cdot\mathbf{S}_{i+1}+J_2\sum_i
  \mathbf{S}_i\cdot\mathbf{S}_{i+2},\ J_1,J_2>0\ , 
\end{equation} 
the situation is similar. For the calculations presented in this paper,
we set $J_1 \equiv 1$ and vary the remaining parameter $J_2$.  The
disorder point, with minimal correlation length ($\xi\simeq 0$), occurs
at the Majumdar-Ghosh~\cite{majumdar_antiferromagnetic_1970} (MG) point,
$J_2^d=J_1/2$, and the Lifshitz point at
$J_2^L=0.52036(6)J_1$.~\cite{bursill_numerical_1995} 
\begin{figure}[htb]
  \begin{center}
    \includegraphics[width =0.47 \textwidth]{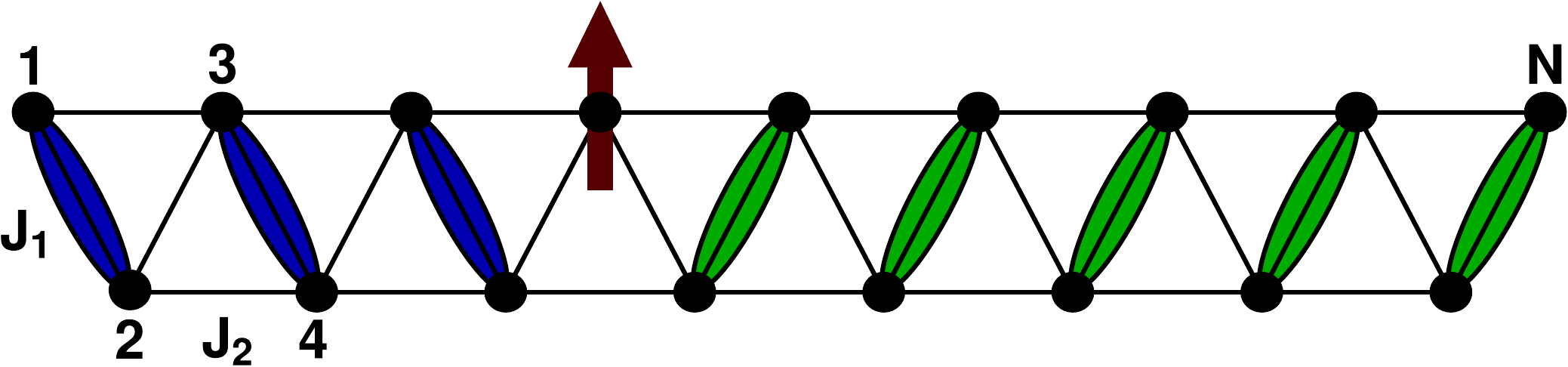}
  \end{center}
  \caption{ (Color online.) The odd length $J_1$-$J_2$ chain. Two
  different dimerization patterns are separated by a soliton.}
  \label{fighamiltonian}
\end{figure}

At the disorder points, the ground states of these two quantum spin
models share important features: the system is gapped and an exact wave
function is known. For both,  the momentum of the lowest excitations
changes at a distinct point.  Yet, there are also important differences
between the two systems.  While the $S=1$ VBS state remains an exact
state for a chain with an {\it odd} number of spins, this is {\it not}
the case for the $S=1/2$ $J_1$-$J_2$ chain at the MG point where no
analytical expression for the odd length
ground state wave function is known.  Moreover, the odd length
$J_1$-$J_2$ chain is gapless in the thermodynamic limit within the
$S_{Tot}=1/2$ subspace, and a large spin gap exists. The onset of
incommensurability effects for odd length chains must then be quite
distinct from the onset in even length chains.  Here, we show that this
is indeed the cases. While the disorder point remains unchanged, the
nature of the Lifshitz point, $J_2^L\simeq0.538 J_1$, is rather different.
At $J_2^L$, a sequence of level crossings starts, changing the
ground state momentum away from $k=\pi/2$. Although the correlation
length remains small, the change in ground state momentum induces
pronounced oscillations directly in the on-site magnetization  as well
as the entanglement entropy.  The modulations in the on-site
magnetization are potentially observable in experiments.  This scenario
is reminiscent of a real Lifshitz
transition~\cite{Lifshitz_1960,Hornreich_1975,Blanter_1994,Yamaji_2006}
in which the ground state becomes modulated. The scaling of the
entanglement entropy at Lifshitz transitions recently has been the
subject of interest.~\cite{Fradkin_09,Rodney_2012}

The $S=1/2$ $J_1$-$J_2$ antiferromagnetic (AF) spin chain is one of the
simplest frustrated Heisenberg spin models, but it has a rich  phase
diagram. The system undergoes a
transition~\cite{haldane_spontaneous_1982-1} from a gapless Luttinger
liquid to a dimerized phase at a critical value of $J^c_2=0.241167
J_1$.~\cite{tonegawa_ground-state_1987,okamoto_fluid-dimer_1992,Eggert_1996}
For {\it even } length chains the ground state wave function at the MG
point $J_2=J_1/2$ is known to be formed by nearest neighbor
dimers.~\cite{majumdar_nextnearestneighbor_1969,majumdar_antiferromagnetic_1970,broek_exact_1980}
It is two-fold degenerate, corresponding to the two possible nearest
neighbor dimerization patterns, indicated in Fig.~\ref{fighamiltonian}.
As is evident from Fig.~\ref{fighamiltonian}, an unpaired spin, a
soliton,~\cite{shastry_excitation_1981,caspers_majumdar-ghosh_1984} can
act as a 'domain wall' and separate regions of different dimerization
patterns. In the Luttinger liquid phase unpaired spins are more commonly
called spinons since they do not act as domain walls. Spin excitations
in the even length chain correspond to introducing two solitons and it
is known~\cite{soerensen_soliton_1998} that in the vicinity of the MG
point the solitons do {\it not} bind and a large spin gap of
$\Delta=2\Delta_{sol}$ (at the MG point
$\Delta_{sol}/J_1=0.1170(2)$)~\cite{soerensen_soliton_1998} exists.  The
spin gap for even length chains is known to remain
sizable~\cite{Chitra_1995, white_dimerization_1996} beyond $J_2^L$. The
presence of a large soliton mass, $\Delta_{sol}$, renders variational
calculations based on a reduced Hilbert space consisting of soliton
states very precise;~\cite{shastry_excitation_1981,caspers_majumdar-ghosh_1984} a fact that
we shall exploit here.

In contrast, for {\it odd} length chains it is not possible for the
chain to be fully dimerized and the ground state wave function is not known
for any value of $J_2$. An $S=1/2$ soliton that effectively behaves as a
free particle~\cite{uhrig_unified_1999} is always present in the
ground state and gives rise to gapless excitations.  Depending on the
quantity in question, odd and even length chains can show very different
behavior.  Under open boundary conditions (OBC), this has for example
already been seen in the on-site magnetization,
\cite{soerensen_soliton_1998} the entanglement entropy
\cite{soerensen_quantum_2007,Affleck_2009} and the negativity.
\cite{deschner_impurity_2011} As mentioned, here we focus 
on odd length chains.

While it is possible to perform highly precise density matrix
renormalization group (DMRG) calculations well beyond the onset of
incommensurability for even
chains,~\cite{white_dimerization_1996,Chitra_1995} the sequence of level
crossings that we encounter for odd length chains for $J_2>J_2^L$
significantly restrains the usefulness of the DMRG technique in a large
region of parameter space for $J_2>J_2^L$. Fortunately, using the
picture of Shastry and Sutherland,~\cite{shastry_excitation_1981} it is
possible to quite efficiently perform very precise variational
calculations for both open and periodic boundary conditions (PBC). Here,
we mainly present results of such variational calculations
and supplement them with DMRG-results.

A number of spin-Peierls compounds, which to some extent realize the
$J_1$-$J_2$ spin chain, have been identified. One of the most well known
is CuGeO$_3$.~\cite{Hase_1993} In these materials impurities
often cut the chains at random points. Therefore both odd and even
length chains are present. A particular point of focus has been the
study of $S=1/2$
solitons~\cite{FagotRevurat_1996,FagotRevurat_1997,Horvatic_1999,Uhrig_solitons_1999}
in these systems. Thus, our results might be directly verifiable if
materials with sufficiently large $J_2>J_2^d$ can be found.

The outline of the paper is as follows. In section~\ref{sec:method} the
variational approach is described.  Section~\ref{sec:incom} begins with
a presentation of our DMRG results for the correlation functions,
correlation lengths and the structure factor.  In section~\ref{subsec:pbc}
we discuss our variational results for the $J_1$-$J_2$ with periodic
boundary conditions and show the change in ground state momentum
developing at the Lifshitz point. Section~\ref{subsec:obc} contains
variational and DMRG results for the on-site magnetization and level
crossings occurring with open boundary conditions.  Variational and DMRG
results for the entanglement entropy for a range of $J_2$ for odd length
chains (OBC) are presented in section~\ref{sec:entang} and contrasted
with results for {\it even} length chains (OBC).  Finally, 
estimates for the location of the Lifshitz point are presented in
section~\ref{sec:trans}. 

In the following, we shall take $J_1\equiv 1$.  This leaves us with only
one parameter, $J_2$, that governs the properties of the system.

\section{The variational method}
\label{sec:method}
Most of the results presented in this paper were generated using
variational
calculations,~\cite{shastry_excitation_1981,caspers_majumdar-ghosh_1984,Zeng_1995,soerensen_quantum_2007,deschner_impurity_2011}
i.e. the results were obtained by minimizing
the expectation value of the Hamiltonian within a reduced Hilbert-space:
\begin{eqnarray}
  \langle H \rangle = \frac{( \varphi | H \varphi )}{(
  \varphi | \varphi )} \ ,
  \label{eq:var_definition}
\end{eqnarray}
where 
\begin{equation}
  \varphi = \sum c_j \varphi_j  
  \label{eq:phi}
\end{equation}
and the minimization is done with respect to the $c_j$. To get a good
estimate of the true ground state of the system, it is necessary that
the ground state has a sizable projection onto the subspace one
diagonalizes in. The quality of the result of a variational calculation
thus depends very strongly on the choice of subspace.  Often one has to
rely on physical insight and intuition to choose well.   For the \JJ, 
which we consider, the selection of an appropriate subspace is
straight-forward as long as one stays in the dimerized phase. In
contrast,  in the Luttinger liquid phase, selecting an appropriate
subspace seems intractable.

\begin{figure}[htb]
  \centering
  \subfloat[]{
  \includegraphics[width = 0.25 \textwidth]{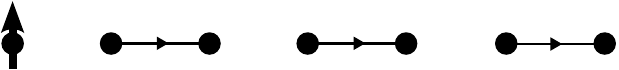}
  }

  \subfloat[]{
  \includegraphics[width = 0.25 \textwidth]{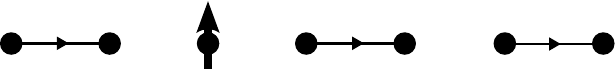}
  }
  \caption{Two variational states used in the calculations
  within $R_0$ for a chain with an odd number of sites. The arrows
  between sites are used to fix the phase of the dimers (see main
  text).}
  \label{figtrialstatesodd1}
\end{figure}

The first variational calculations on the \JJ \ were done in a space
that we in the following shall call $R_0$.
\cite{shastry_excitation_1981,caspers_majumdar-ghosh_1984}  It is
spanned by the states in which there are domains that have one of the
two ground state configurations of the MG-chain and which are separated
by one soliton.  Examples can be seen in Figs.~\ref{fighamiltonian} and
\ref{figtrialstatesodd1}.  The arrows in Fig.~\ref{figtrialstatesodd1}
serve to fix the phase of the dimers that make up the ground state.  Our
convention is such that if the arrow goes from site $i$ to site $j$
the spins are in the state:
$\frac{1}{\sqrt{2}}( | \uparrow\rangle_i |\downarrow \rangle_j - |
\downarrow \rangle_i |\uparrow \rangle_j ) $.
For a chain with an odd number of sites, a set of single soliton states
can be generated by leaving the chain maximally dimerized 
and taking the remaining site to be in the $S_z$$=$$1 /
2 $-state. For a chain with open boundary-conditions, the soliton can
only reside on every second site.  The dimension of this
variational subspace is then $D$$=$$(N + 1) / 2 $. We use $N$ to denote
the length of the chain.  For calculations on odd length chains with
periodic boundary conditions it is necessary to allow a nearest neighbor
dimer across the boundary and to let the soliton cross the boundary by
going from site $N$ to site 2. In this case $R_0$ has dimension $N$ and
incorporates states with the soliton at every site with the remaining
spins paired in nearest neighbor dimers. (For odd $N$ and PBC it becomes
difficult to distinguish the 2 dimerization patterns since they twist
into each other at the boundary. Still, the soliton clearly denotes a
'domain wall' between the two patterns).

\begin{figure}[htb]
  \centering
  \subfloat[]{
  \includegraphics[width = 0.25 \textwidth]{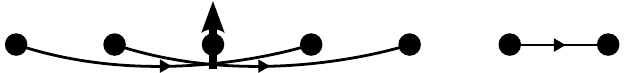}
  }

  \subfloat[]{
  \includegraphics[width = 0.25 \textwidth]{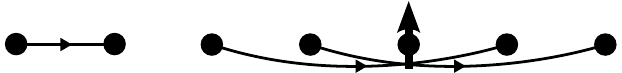}
  }
  \caption{Two variational states used in the calculations
  within $R_1$ for a chain with an odd number of sites. The arrows
  between sites are used
  to fix the phase of the dimers (see main text).}
  \label{figdiffenltrialstates}
\end{figure}

\begin{figure}[htb]
  \centering
  \subfloat[]{
  \includegraphics[width = 0.25 \textwidth]{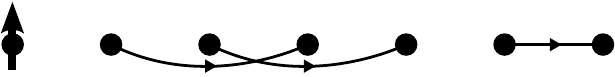}
  }

  \subfloat[]{
  \includegraphics[width = 0.25 \textwidth]{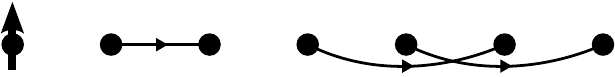}
  }

  \subfloat[]{
  \includegraphics[width = 0.25 \textwidth]{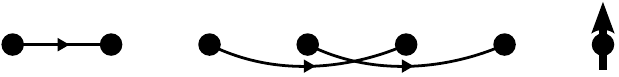}
  }
  \caption{Three states that are generated by acting with the Hamiltonian
  onto $R_0$. The arrows between sites are used to fix the phase of the
  dimers (see main text).}
  \label{figtrialstatesodddist}
\end{figure}

To improve upon $R_0$, it is natural to act with the Hamiltonian onto
the space as doing this repeatedly generates a space that contains the
ground state if the starting space had any overlap with and all
symmetries of the ground state.  It was shown that acting onto $R_0$
with the Hamiltonian only once is at the MG point already enough to make
the calculation almost exact.\cite{caspers_majumdar-ghosh_1984} For the
\JJ \ with $J_2 \neq 0.5J_1$ the linearly independent states generated by
acting with the Hamiltonian onto $R_0$ fall into three classes, each of
which corresponds to a variational subspace:%
\begin{itemize}
  \item {\bf The variational space $R_0$}: Spanned by the states that are in $R_0$.
  \item {\bf The variational space $R_1$}: Spanned by states in which sites to the left and right of
    the soliton are connected by a valence bond. Pictorial
    representations of example-states are shown in
    Fig.~\ref{figdiffenltrialstates}. 
  \item {\bf The variational space $R_2$}: Spanned by states in which two neighboring sites are in a
    valence bond with their next-nearest neighbor. These states are
    generated by the action of the nearest-neighbor-terms and the
    next-nearest neighbor terms in the Hamiltonian on adjacent dimers in
    the states in $R_0$. Pictorial representations of example-states are
    shown in Fig.~\ref{figtrialstatesodddist}.  In the case of the
    MG-chain, $J_1$ and $J_2$ are balanced in such a way that these
    states are not generated because they occur with a weight of $2 J_2
    - J_1$. 
\end{itemize}

The number of states in $R_0$ and $R_1$ scales linearly with the size of
the chain, whereas the number of states in $R_2$ scales quadratically.
Due to computational cost we have thus not found it practical to use the
union of the three as the variational subspace for chains longer than
101 sites.  We performed calculations using the union of $R_0$ and
$R_1$ (in the following called $Z_s$) for chains up to $1001$ and the
union of all three (in the following called $Z_b$)
for a chain of $101$ sites. In this way we could go to long chains and
also check the validity through the comparison at $N=101$. We found that
while there were small quantitative differences between calculations done in
$Z_s$ and $Z_b$ the overall qualitative features of the results where
the same.
Therefore, we chose to use $Z_s$, the union of $R_0$ and $R_1$, or 
just $R_0$ for the variational calculations shown in this paper.

All the states in these spaces have $S_z^{Tot} := \sum_i S_z^{i} = 1/ 2$. We could equally
well have worked in the  $S_z^{Tot}=-1/2$ space.  States of higher total
spin are of little importance to the low-energy physics since they
contain more solitons and are thus gapped by at least twice the soliton
mass $\Delta_{sol}$.  Since $\Delta_{sol}$ is
sizable~\cite{soerensen_soliton_1998} in the regime of our study, such
states can be disregarded for both odd and even $N$.

A variational description of a chain with an even number of sites can be done
along the same lines.  Again, the states are chosen in order to leave all but
two spins in the favored dimerized state.  In this way one can gain insight
into the low-energy singlet- as well as triplet-excitations by choosing the two
spins to be in singlet- or the triplet-states respectively.

If one considers a subspace with  an orthogonal basis, one can just
diagonalize the Hamiltonian.  While an easy way to orthogonalize $R_0$
is known,~\cite{uhrig_unified_1999} this is generally not true for other
subspaces. Importantly, for  $R_1$ no such method is known.  We thus
have to solve the generalized eigenvalue-problem given by
\begin{align} 
  \mathfrak{H} \; \vec{\varphi} = \lambda \; \mathfrak{B} \vec{\varphi} \ , 
\end{align}
where 
$\mathfrak{H}_{ij}:= (\varphi_i| H \varphi_j)$ 
and
$\mathfrak{B}_{ij} := (\varphi_i| \varphi_j)$.  
Such generalized-eigenvalue-problems can be solved numerically by
standard routines.  We calculate $\mathfrak{H}$ and $\mathfrak{B}$ by
evaluating their defining expressions. This is possible because for
valence-bond-states the action of $H$ on them as well as the overlap
between them can straightforwardly be calculated in an automated manner.
How to do all other calculations necessary to get the results presented
in this paper has already been described in an earlier
publication.~\cite{deschner_impurity_2011}  We took the coefficients
$c_j$ in Eq.~(\ref{eq:phi}) to be real. Also for PBC, the resulting wave function is
not an eigenstate of the translational operator which would have
required the use of complex $c_j$'s. Effectively we obtain states that
are linear combinations of translationally invariant states with $k$ and
$-k$, degenerate in energy. While this has no bearing on the obtained
energies, it affects real-space quantities like the on-site
magnetization and entanglement which cannot be translational invariant.

\section{The incommensurate behavior}
\label{sec:incom}
\begin{figure}[th]
  \begin{center}
    \includegraphics[]{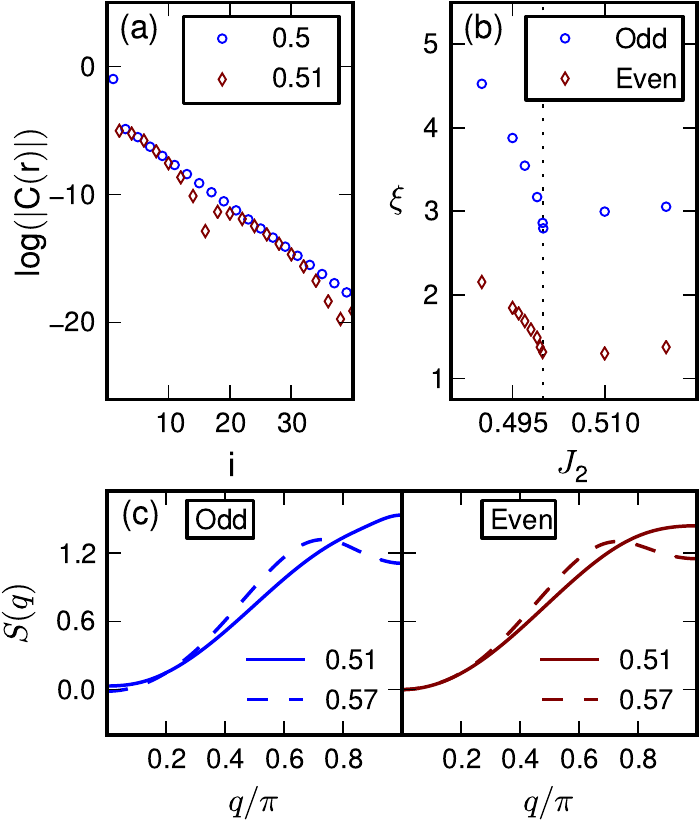}
  \end{center}
  \caption{(Color online.) (a) The spin-spin correlation function
  $C(r)=\langle S_iS_{i+r} \rangle$ for a chain with 201 sites and
  $J_2=0.5, 0.51$. The correlation-length (b) and the static
  structure factor (c) for chains with an odd (201) or an even (200)
  number of sites. The data was obtained with DMRG.  Open boundary
  conditions were employed and $m=256$ states kept. For both, odd  and even length
  chains, the correlation-length displays a minimum at the MG point.
  The static structure factors of odd and even are shown for $J_2 =
  0.51$ and $J_2  = 0.57$. The maximum remains at $q=\pi$ until the
  Lifshitz point is crossed (not shown in figure). 
  }
  \label{figcorre}
\end{figure}

Previous numerical studies of disorder points in
$S=1/2$~\cite{tonegawa_ground-state_1987,Nomura_1993,Nomura_1994,bursill_numerical_1995,Chitra_1995,white_dimerization_1996}
and
$S=1$~\cite{schollwoeck_onset_1996,Kolezhuk_1996,Roth_1998,Polizzi_1998,Golinelli_1999}
quantum spin chains have concentrated on the behavior of even length
chains. For the $S=1/2$ $J_1$-$J_2$ chain it has been shown that the
disorder point of this $1$-dimensional quantum system can be understood
as a $1+1$ dimensional classical disorder point. In particular, it was
shown~\cite{Garel_1986, Fath_2000, Nomura_2002} that in the
``commensurate'' region of the phase diagram the correlation function
behaves asymptotically, with $r\gg \xi$, as
\begin{equation}
  \langle S_iS_{i+r}\rangle\sim (-1)^r\frac{e^{-r/\xi}}{\sqrt{r}},
  \label{eq:corC}
\end{equation}
and in the ``incommensurate'' region of the phase diagram as
\begin{equation}
  \langle S_iS_{i+r}\rangle\sim (-1)^r\frac{e^{-r/\xi}}{\sqrt{r}}\cos\left[(q-\pi) r+\phi\right].
  \label{eq:corIC}
\end{equation}
Here, $q$ is the wave vector of the incommensurate correlations and $\phi$ a phase shift.
However, right at the disorder point separating commensurate and incommensurate correlations
the correlation function is asymptotically purely exponential:
\begin{equation}
  \langle S_iS_{i+r}\rangle\sim (-1)^r e^{-r/\xi}.
  \label{eq:corD}
\end{equation}
For these quantum spin models it appears that this purely exponential
behavior is in part connected to the fact that the ground state is an
exact nearest neighbor dimer state. Interestingly, as we shall see, the
correlation functions at the MG point for odd length chains display the
same behavior in the absence of a unique nearest neighbor dimer
ground state.  Furthermore, it is known~\cite{Garel_1986,Fath_2000} that
as the disorder point is approached from the commensurate side, the
derivative of the correlation length with respect to the driving
coupling becomes infinite, while it is finite on the incommensurate
side.  It is also known that the disorder point has special degeneracies
that are exact for any system size $N$. For instance, for the
$J_1$-$J_2$ chain with periodic boundary conditions and an even length,
the two dimerization patterns are degenerate at the disorder point while
their symmetric and antisymmetric combinations are split with an
exponentially small gap away from this point.

We now present our results for the incommensurate effects in odd length
$S=1/2$ $J_1$-$J_2$ chains.  Our first point of focus is the location of
the disorder point. As stressed above, when $N$ is odd, the nearest
neighbor dimer wave function is {\it not} an exact
solution~\cite{shastry_excitation_1981} and there are also no special
degeneracies. There is therefore no reason to expect that the behavior
of the correlation length at the MG point is in any way unique. However,
as we shall see, this is indeed the case. DMRG results for $C(r)=\langle
S_i S_{i+r}\rangle$ for an open chain with $N=201$ are shown in
Fig.~\ref{figcorre}(a) for $J_2=0.5$ and $0.51$.  The correlation
function follows a {\it purely} exponential decay at the MG point,
$J_2=0.5$, with a finite correlation length:
\begin{equation}
  \xi_{\mathrm{MG}}\sim 2.8 \ \ (\mathrm{odd} \ N).
\end{equation}
Distant spins in odd chains are correlated even at the MG point, because
the soliton is present in the chain. The correlations can be thought of
as correlations in the soliton wave function.
Secondly, as can be seen in Fig.~\ref{figcorre}(a), incommensurate
correlations are clearly present for $J_2=0.51$. They were present in
every calculation we performed with $J_2>0.5$.  We conclude that the
disorder point remains at $J_2=J_1/2$ albeit with a finite correlation
length compared to the case of even $N$ where the correlation length is
nominally zero.

The precise behavior of the correlation length around the disorder point
$J_2=1/2$ appears to have been studied neither for even length nor for
odd length chains. Results for larger $J_2>0.6$ are available for even
$N$.~\cite{white_dimerization_1996} By fitting DMRG results for chains
of $200$ and $201$ sites to the forms Eqs.~(\ref{eq:corC}), and
(\ref{eq:corIC}) we have determined $\xi$ as a function of $J_2$ for
both even and odd $N$ (see Fig.~\ref{figcorre}(b)). The results for the
even and the odd length chain are remarkably similar.  At the disorder
point there is a discontinuity in the slope of $\xi$ and on the
commensurate side the slope of $\xi$ approaches $-\infty$. We found that
close to the disorder point in the commensurate region combined forms
like $|C(r)|\sim C \exp(-r/\xi_C)/\sqrt{r} + D\exp(-r/\xi_D)$ with
$\xi_C>\xi_D$ fit the data better than the single forms
Eq.~(\ref{eq:corC}), and (\ref{eq:corIC}), because the dominant
short-ranged correlations change at the disorder point. The results
presented in Fig.~\ref{figcorre} do not use such combined forms. We also
note that for odd $N$ and for a range of $J_2>0.538$ it becomes very
difficult to obtain reliable DMRG results due to the appearance of many
almost degenerate states.

The structure factor for even chains has been studied in some detail
previously,~\cite{tonegawa_ground-state_1987,bursill_numerical_1995} and
the Lifshitz point has been located,
$J_2^L=0.52036(6)$.~\cite{bursill_numerical_1995}  Our DMRG results
are shown in Fig.~\ref{figcorre}(c).  In agreement with previous studies
for even $N$, we observe that the maximum in the structure factor
remains at $q=\pi$ for $J_2=0.51$ but has clearly moved away from $\pi$
at $J_2=0.57$. This is clearly also the case for odd $N$. Due to the
above mentioned difficulties in obtaining reliable DMRG results for odd
$N$ and $J_2>0.538$ we have not been able to determine the precise point
where the peak in the structure factor is displaced from $q=\pi$.  Using
the variational techniques outlined above it is possible to understand
in detail what happens close to $J_2\simeq0.538$.  
\begin{figure}[ht]
  \includegraphics[]{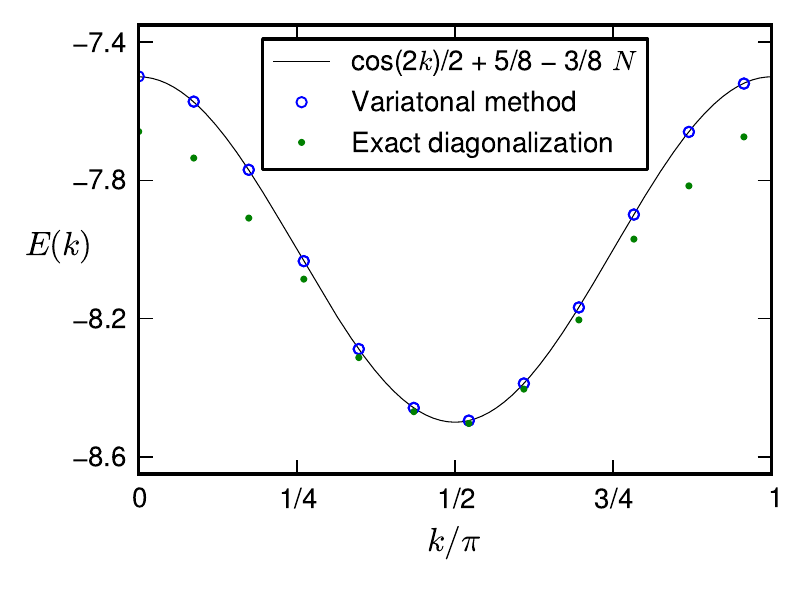}
  \caption{(Color online.)
  Comparison of the estimate by Shastry and
  Sutherland~\cite{shastry_excitation_1981}, the variational method and
  exact diagonalization (ED) data.~\cite{soerensen_soliton_1998}  Data
  was taken for a chain with 23 sites and $J_2$=$0.5$. Only the lowest
  energy of a spin $1/2$ excitation for every momentum is shown.  The
  deviations to ED occur at higher energies where the dispersive mode
  enters the continuum and the variational calculation is only of
  limited value.  
  }
  \label{figdmrgdisp}
\end{figure}
\subsection{Variational Results in Periodic Boundary Conditions}
\label{subsec:pbc}
We now turn to a discussion of our variational results obtained using
the method outlined in section~\ref{sec:method}. We begin by focusing on
the case of odd length chains and periodic boundary conditions. The case
of open boundary conditions will be the subject of the next subsection.
The results shown in this subsection were obtained
using the space $R_0$ (see section~\ref{sec:method}), consisting of all
single soliton states with $S_z^{Tot}=1/2$.

At the MG point the spectrum of the \JJ \ has been studied extensively.
The feature that is most important to us is the low-lying dispersive
line that is well separated from the
continuum\cite{soerensen_soliton_1998} and roughly follows a cosine as
found in previous variational
studies.~\cite{shastry_excitation_1981,Arovas_1992}  
\begin{equation}
  E(k)=\frac{1}{8}\left(5+4\cos(2k)-3N\right).
\end{equation}
Our variational method reproduces this estimate and agrees well with the
low-energy data of an exact diagonalization of a chain of 23 sites (see
Fig.~\ref{figdmrgdisp}).  It may be surprising that the minimum of the
dispersion relation is not at $k = \pi$ but at $k=\pi / 2$. This is a
natural consequence of the effective doubling of the unit cell that
occurs because the action of the Hamiltonian displaces the soliton by
{\it two} sites.

\begin{figure}[ht]
  \begin{center}
    \includegraphics[]{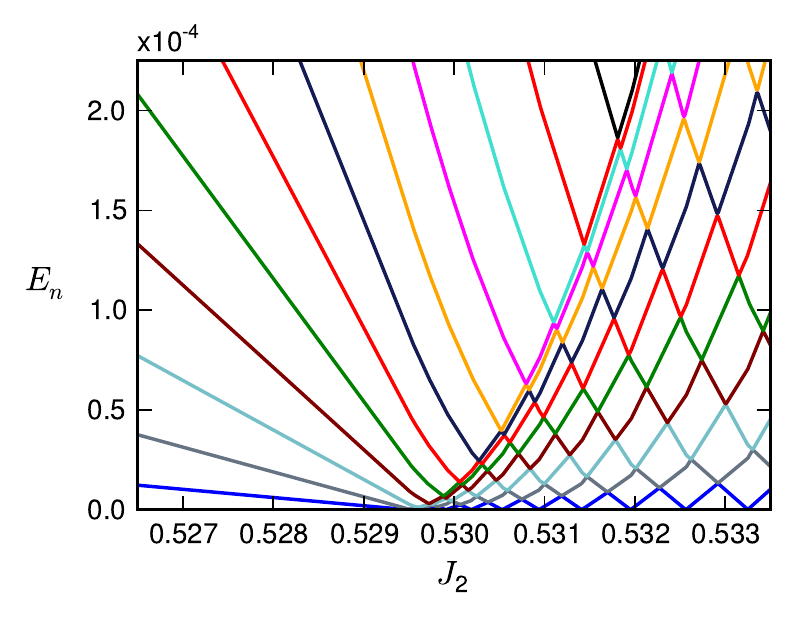}
  \end{center}
  \caption{(Color online.) The excitation spectrum with periodic
  boundary conditions in the variational subspace with $S_z^{Tot}=1/2$.
  The energies of the first few excited states are shown. The
  ground state energy was set to zero.  The energies of the excited
  states approaches the ground state energy until, at $J_2 = J_2^L$,
  level-crossings start to occur. All energy-levels are doubly
  degenerate. The data was taken for a chain with 401 sites.
  }
  \label{figspecpbc}
\end{figure}

One of the strengths of the variational method is that within the limits
of the approximation it is possible to easily access not only the
ground state but also the entire energy spectrum within the variational
subspace of $S_z^{Tot}=1/2$ states.  Computing the spectrum through the
transition region reveals very surprising behavior (see
Fig.~\ref{figspecpbc}). All the states are two-fold degenerate
corresponding to the energetically degenerate $k$ and $-k$.  As one
approaches the transition, the excited states linearly move closer
and closer to the ground state. At the Lifshitz point $J_2^L
\approx 0.53$ the energy of the first excited state crosses the
ground state energy. This level crossing marks the first shift in the
ground state momentum and is followed by a series of other level
crossings at larger $J_2$ that further shift the ground state momentum.
Clearly, the presence of the many adjacent level crossings
hinders the effectiveness of DMRG calculations.

This is in stark contrast to the spectrum of even length chains: The
ground state of even length chains is exactly two fold degenerate at the
MG point for any $N$ whereas for larger $J_2$ the symmetric and
antisymmetric combinations are split with an exponentially small gap in
$N$.  The excited states are separated from these two states by a large
gap of approximately $2\Delta_{sol}$. This gap persists throughout the
transition region and no level crossings are
observed.~\cite{tonegawa_ground-state_1987, white_dimerization_1996}
\begin{figure}[ht]
  \begin{center}
    \includegraphics[]{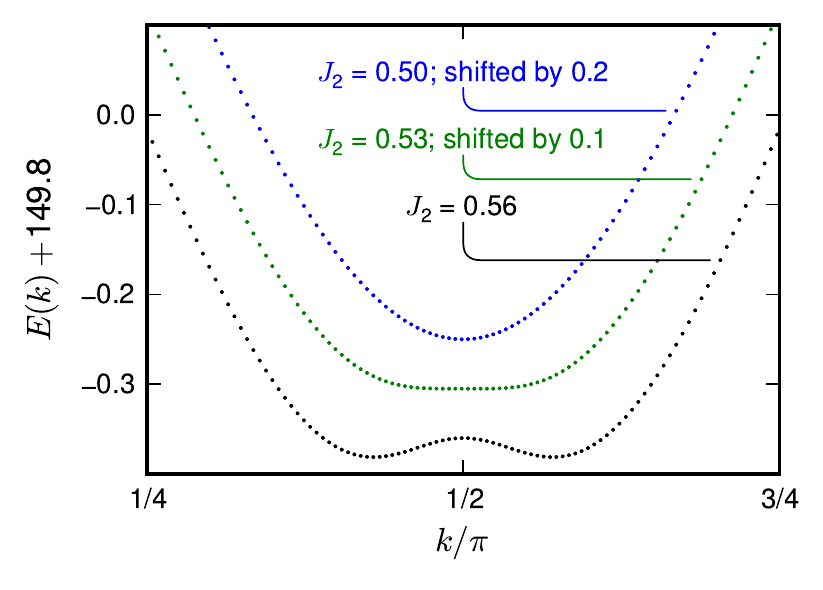}
  \end{center}
  \caption{(Color online.) Dispersion relation for varying $J_2$. The data was taken for
  a chain with 401 sites. To avoid cluttering, the dispersion relations
  for the smaller two $J_2$ were shifted. As $J_2$ is increased, the
  minimum first flattens and then turns into a local maximum.
  }
  \label{figchadisp}
\end{figure}

It is very instructive to look at how the dispersion relation in
Fig.~\ref{figdmrgdisp} evolves with $J_2$. As can be seen in
Fig.~\ref{figchadisp}, the dispersion relation changes its shape when
$J_2$ is increased.  The minimum at $k=\pi/2$ first becomes flat very
close to $J_2=0.53$ and then becomes a local maximum. In the process two
minima  are created, which move away from $k=\pi/2$ with increasing
$J_2$. The ground state momentum is then clearly changing away from
$k=\pi/2$ beyond $J_2=0.53$ and we may identify the point where this
happens with a real Lifshitz
transition~\cite{Hornreich_1975,Hornreich_1978,Hornreich_1980,Michelson_1977a,Michelson_1977b,Michelson_1977c}
as opposed to the corresponding point in the $S=1$ bilinear biquadratic
chain where the ground state momentum remains unchanged and the shift is
in the excited magnon dispersion.  Due to the shift in the ground state
momentum we conclude that the maximum of the structure factor will shift
away from $k=\pi$ at the same point. This is consistent with the data in
Fig.~\ref{figcorre}. We therefore in the following refer
to this point as the Lifshitz point, $J_2^L$. %
\begin{figure}[th]
  \begin{center}
    \includegraphics[]{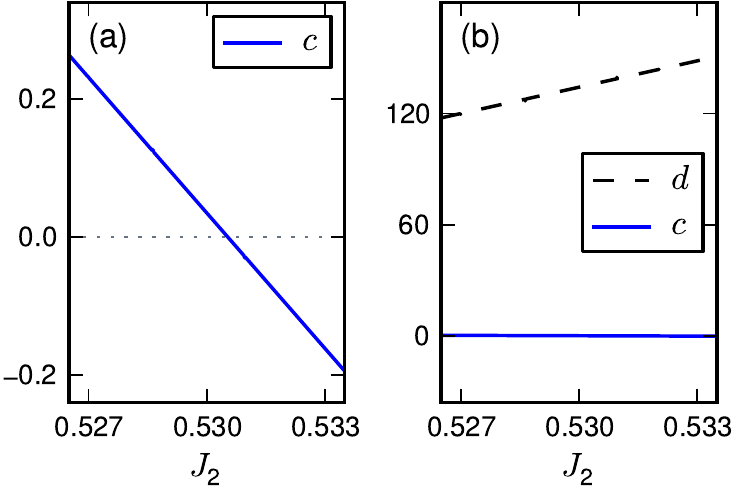}
  \end{center}
  \caption{(Color online.) The second-order coefficient of the
  dispersion relation $c$ changes sign ((a), (b)) while the forth-order
  coefficient $d$ stays positive (b). The data are results off a fit to
  data of the kind shown in Fig.~\ref{figchadisp}. The dotted horizontal
  line in (a) separates positive from negative values.
  }
  \label{figdiscoeff}
\end{figure}
The precise behavior of the dispersion relation close to $J_2^L$ is
analyzed in Fig.~\ref{figdiscoeff}. 

For a range of $J_2$, we fitted the dispersion relation to
the form~\cite{Golinelli_1999}
\begin{equation}
  E(k)=E(k_0)+\frac{c}{2}(k-k_0)^2+\frac{d}{24}(k-k_0)^4
\end{equation}
and confirmed that the second-order coefficient $c$ changes its sign  at
a $J_2$ close to 0.53 while the fourth order coefficient $d$ stays
positive (see Fig.~\ref{figdiscoeff}).  This behavior is typical of a
Lifshitz transition and if the coefficient $c=v^2/\Delta_{sol}$ is
associated with a velocity, $v$, the Lifshitz point signals the
vanishing of this velocity.~\cite{Golinelli_1999}

The variational calculations with periodic boundary conditions presented
in this section were limited to the subspace $R_0$ described in
section~\ref{sec:method}. This basis only includes nearest neighbor
valence bonds and it is quite noteworthy that the physics of the
Lifshitz point along with the associated level crossings are captured
within this simple basis set. However, as we discuss in
section~\ref{sec:trans} we do not expect the precise location of the
Lifshitz point to be accurately determined within $R_0$. 

\subsection{Open boundary conditions}
\label{subsec:obc}
In materials that realize the $J_1$-$J_2$ spin chain impurities are
always present. They often act as non-magnetic impurities effectively
breaking the linear chains into finite segments. The use of open
boundary conditions is therefore closer to the experimental situation
than the use of periodic boundary conditions. Furthermore, it is natural
to expect half of the chain segments to have an odd number of sites.  In
this subsection we therefore focus on odd length chains with open
boundary conditions. In particular, we describe the change that
switching from periodic to open boundary conditions causes.  The
variational results shown in this subsection were obtained using the
space $Z_s$ (see section~\ref{sec:method}).
\begin{figure}[ht]
  \begin{center}
    \includegraphics[]{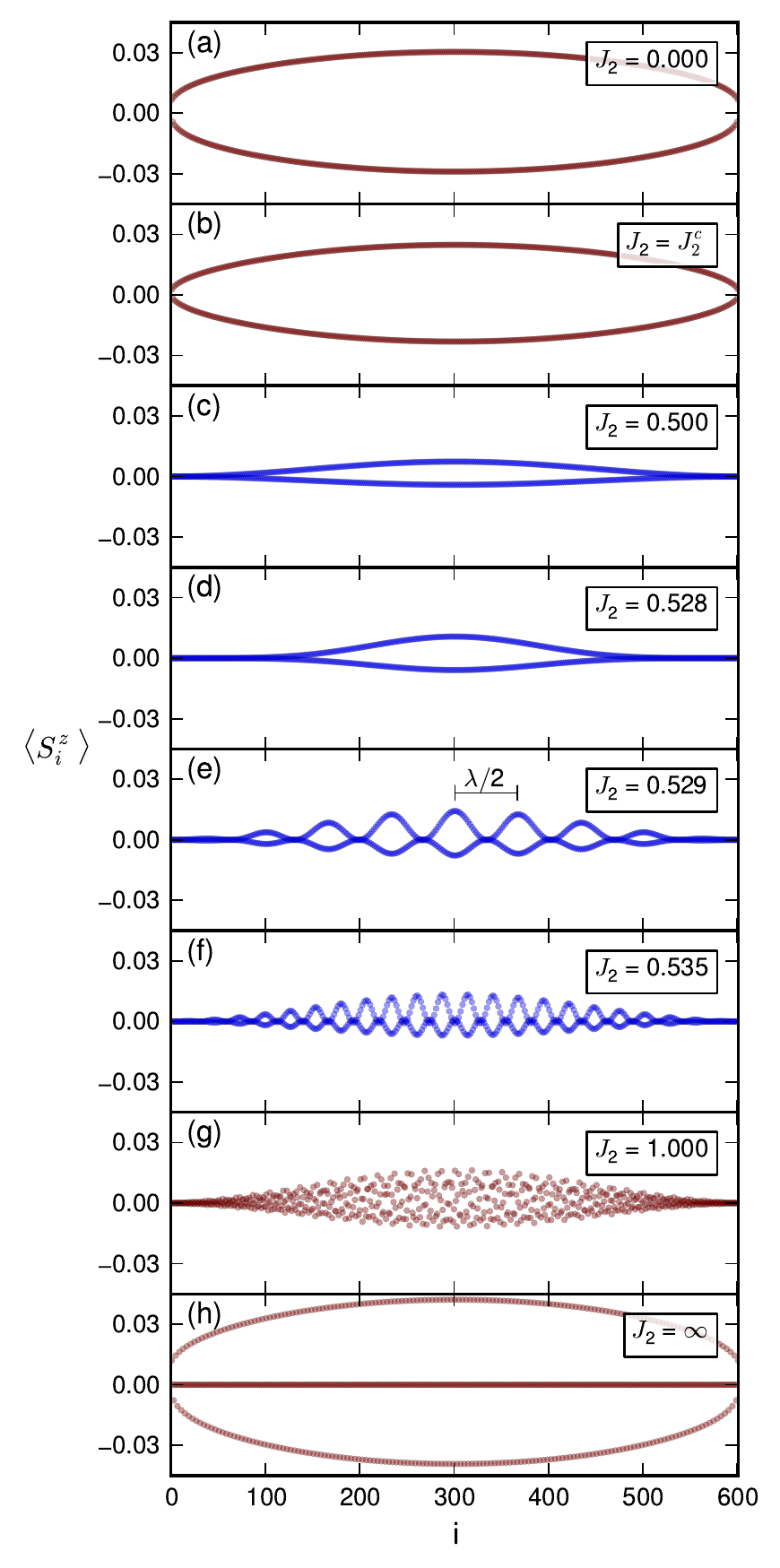}
  \end{center}
  \caption{(Color online.) The on-site magnetization at a range of
  values of $J_2$ for $N=601$ sites. Additional structure appears beyond
  $J_2^L$.  For $J_2 = 0$, $J_2^c$, $1$ and $\infty$ DMRG-data (obtained with
  $m=256$ states kept) are shown (red). For the remaining $J_2$
  variational calculations are shown (blue). }
  \label{figonsiteincm}
\end{figure}

One quantity that very directly shows the qualitative difference between
PBC and OBC is the ground state on-site or local magnetization, $\langle
S^z_i\rangle$, which is of importance to for instance NMR
measurements.~\cite{FagotRevurat_1996,FagotRevurat_1997}
Figure~\ref{figonsiteincm} shows the
on-site magnetization at 8 different values of the frustrating
interaction $J_2$ between $J_2=0$ and $J_2=\infty$ in a chain of 601
sites. The figures~\ref{figonsiteincm}(c)-(f) show variational
calculations through the Lifshitz point $J_2^L$ where DMRG calculations
are less effective, the remaining results (a),(b),(g) and (h) are
obtained with DMRG.

{\it The Luttinger liquid phase ($J_2\le J_2^c$):} The transition to the
dimerized phase occurs at $J_2^c$, see Fig.~\ref{figonsiteincm}(b). At
this point, as well as throughout the Luttinger liquid phase
($J_2<J_2^c$), the on-site magnetization agrees very well with the
prediction for the on-site magnetization in the ground state with
$S_z^{Tot}=1/2$ from conformal field theory:~\cite{Eggert_2002}
\begin{equation}
  \langle S^z_i\rangle=C(-1)^i\sqrt{\frac{\pi}{2N}\sin\left(\frac{\pi i}{N}\right)}+\frac{1}{2N},
\end{equation}
where $C$ is a constant. In this phase $\langle S^z_i\rangle$ increases
with  the characteristic behavior $\langle S^z_i\rangle\sim\sqrt{i}$ for
small $i$ close to the boundary.

{\it Dimerized phase with $J_2<J_2^L$:} Once the dimerized phase is
entered $\langle S^z_i\rangle$ is drastically altered. The on-site
magnetization roughly follows the behavior of a massive particle in a
box~\cite{soerensen_quantum_2007} with $\langle S^z_i\rangle\sim i^2$
close to the boundary. This behavior is clearly visible at the MG point
(Fig.~\ref{figonsiteincm}(c)). As $J_2$ is increased beyond the MG point
towards the Lifshitz point $J_2^L$ the central peak sharpens
(Fig.~\ref{figonsiteincm}(d)).

{\it 'Incommensurate' phase $J_2>J_2^L$:} At the Lifshitz point there is
another dramatic change in $\langle S^z_i\rangle$: additional maxima
develop and the magnetization is modulated by an oscillating function
(Fig.~\ref{figonsiteincm}(e)). Upon increasing $J_2$ further, more such
maxima form and the wave-length of the modulation decreases (see
figures~\ref{figonsiteincm}(f) and (g)).  
If $J_2$ is fine tuned for a given $N$ it is possible to find a point
where 2 maxima occur in $\langle S^z_i\rangle$, then 3 maxima and so
forth.

It is natural to expect this behavior based on the results for PBC
presented in section~\ref{subsec:pbc}.  The local magnetization is
effectively modulated with the momentum of the ground state. The running
wave found under periodic boundary conditions is converted to a standing
wave under open boundary conditions. Then, as the momentum of the
ground states changes with growing $J_2$, the wave-length of the
modulation shrinks.  Finally, in Fig.~\ref{figonsiteincm}(h) we show
results for $J_2\to\infty$.  In this limit the odd length chain with $N$
sites is split into 2 chains with $(N-1)/2$ and $(N+1)/2$ sites one of
which will have an even number of sites and hence $\langle
S^z_i\rangle\equiv 0$.  The on-site magnetization of the other chain can
be found by calculating $\langle S^z_i\rangle$ for a chain with
$J_2=0$ of the same length. The results shown in
Fig.~\ref{figonsiteincm}(h) were obtained in this way, i.e. from data
for a chain with $N=301$ and $J_2=0$ that was then interspersed with
zeros from the half of the chain that had an even number of sites.
\begin{figure}[t!h]
  \begin{center}
    \includegraphics[]{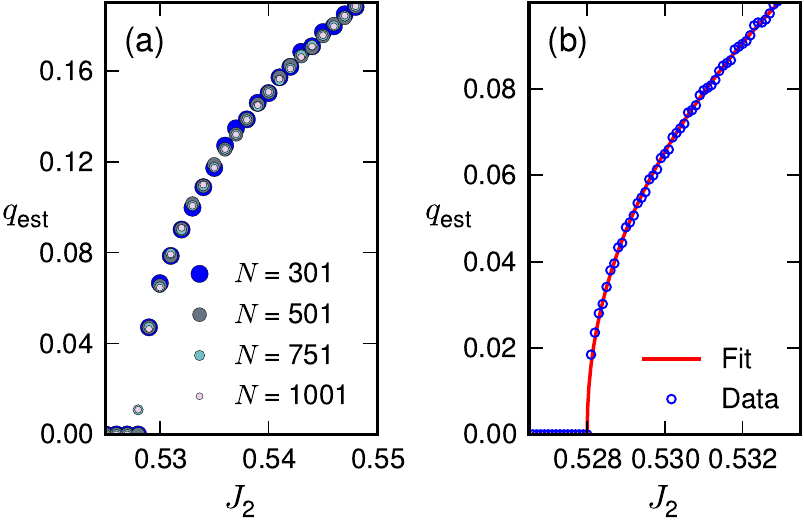}
  \end{center}
  \caption{(Color online.) The wave number $q_\mathrm{est}$ against
  $J_2$.  Data is shown for four chains of different length (a) and with
  the result of a fit $q_\mathrm{est}= 1.2062\ \Theta(J_2 - 0.528)
  (J_2 - 0.528) ^{0.4806}$ for a chain of 601 sites (b). }
  \label{figwavnum}
\end{figure}

To estimate the wave-length of the incommensurate modulation we make use
of the fact that, if our system had translational invariance, the distance
between maxima in the on-site magnetization  would be equal to half of
the wave-length, as indicated in Fig.~\ref{figonsiteincm}(e). Thus, by
calculating the mean distance of the central maxima, we are able to
determine an estimate for the wave-length of the incommensurate
modulation. The inverse of this quantity can then  be used to calculate
the wave number, $q_\mathrm{est}= 2 \pi /
\lambda_\mathrm{est}$.  In Fig.~\ref{figwavnum}(a) we show how
$q_\mathrm{est}$ varies with $J_2$ for four chains whose length ranges
from 301 to 1001 sites. 

Since the incommensurate behavior can only be seen if the wave-length is
shorter than the system, it starts later in smaller chains.  Aside from
small deviations, that can be attributed to finite size effects, the
wave-length does only depend on $J_2$ and not the length of the chain
(see Fig.~\ref{figwavnum}(a)).
In the limit of infinite $J_2$, the next-neighbor interaction $J_1$ can
be neglected and the chain be partitioned into two sub-chains that do
not interact. As mentioned, the \JJ \ in this limit approaches two
uncoupled chains with intra chain coupling $J_2$.  The wave-length of
the incommensurate behavior in this limit reaches its minimum with
$\lambda = 4$ lattice spacings.

For $J_2 > J_2^\mathrm{L}$, one expects the wave number $q$ to behave as
$q \propto (J_2 - J_2^\mathrm{L})^\alpha$, where $0 < \alpha <
1$.~\cite{schollwoeck_onset_1996}  In a study of correlations functions
around the disorder point in the $S=1/ 2$ \JJ \ with an even number of
sites and modified interactions on the edge of the chain, the exponent
was reported to have been calculated to be $\alpha = 1 /
2$.~\cite{nomura_incommensurability_2005}  This is consistent with
calculations on classical Lifshitz points~\cite{Hornreich_1975,
Hornreich_1978} which at the mean field level find $\alpha =1/2$.  In
the present case, where the ground state momentum is changing, one might
also expect corrections to the mean-field value of $\alpha=1/2$ as
described in Ref.~\onlinecite{Hornreich_1975, Hornreich_1978}.

Our calculations indeed confirm that
$q_\mathrm{est}(J_2)$ follows a power-law with exponent smaller than 1
(see Fig.~\ref{figwavnum}( b)).  The line in Fig.~\ref{figwavnum}(b) is
a fit of the three parameter function $f(x) = c_1 \Theta(x - c_2) |x -
c_2|^{\alpha}$, where $\Theta(x)$ is the Heaviside step function, to the
blue data points also shown in the plot.  Using this form we find a
value for the exponent $\alpha = 0.4806$. 
The data in Fig.~\ref{figwavnum} show steplike features.  The cause of
the steps  is the introduction of new maxima: every time a new maximum
appears, $q_\mathrm{est}$ jumps abruptly in order to accommodate the new
maximum and there is a step.  Between the appearance of new maxima, the
maxima that are present move closer together and $q_\mathrm{est}$
increases smoothly. As one increases the system size, this effect 
affects the mean distance between maxima less and thus leads to less
pronounced steps.  Due to the different range of $J_2$-values, the
steplike features explained above are more pronounced in
Fig.~\ref{figwavnum}(b) than in Fig.~\ref{figwavnum}(a).      
Because of the inaccuracies the steplike features introduce to the
fitting procedure, we cannot comment on whether or not the corrections
mentioned above are necessary.
\begin{figure}[ht]
  \centering
  \includegraphics[]{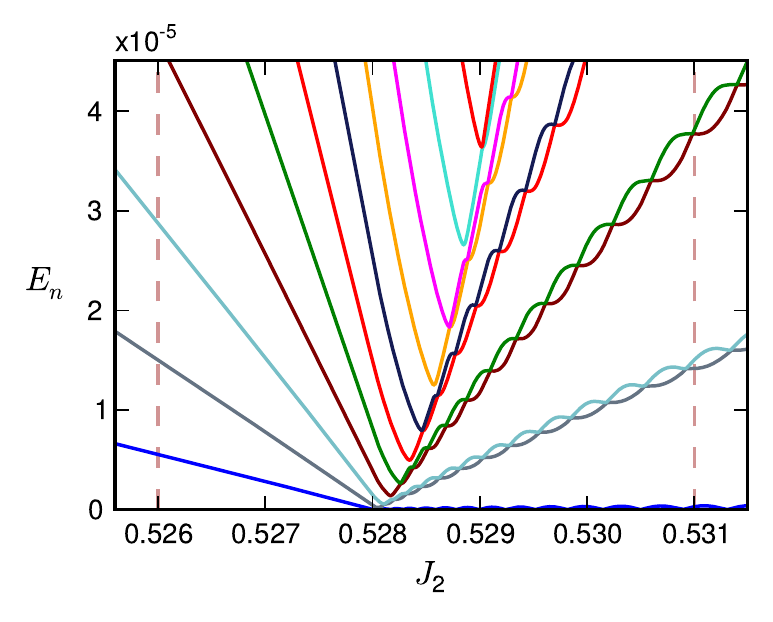}
  \caption{(Color online.) The excitation spectrum with open boundary
  conditions in the variational subspace with $S_z^{Tot}=1/2$.  The 
  energies of the first few excited states are shown. The ground state
  energy was set to zero.  The values at which the scaling with $N$ is
  studied in Fig.~\ref{figscalingwithN} are indicated by dashed vertical
  lines. The data was taken for a chain with 601 sites.
  }
  \label{figspectrum}
\end{figure}

While the on-site magnetization could relatively easily be understood
from the results obtained with PBC, this is not the case for the energy
spectrum.  To the left of the transition ($J_2<J_2^L$), the spectrum for
OBC (shown Fig.~\ref{figspectrum}) looks exactly like the spectrum for
PBC (shown Fig.~\ref{figspecpbc}) --  yet there is an important
difference: the spectrum for OBC is not degenerate.  Introducing the
boundary splits the degenerate states.  On the other side of the
transition ($J_2>J_2^L$), the behavior of the energy of the first
excited state also looks familiar: it hits the ground state energy,
grows, approaches it again and another level-crossing occurs.  Repeated
level-crossings of just the two states follow.  Higher
excitation-levels, however, do not cross many other levels as they do
for PBC. They approach the ground state, then turn around and form a
pair with the state they would have been degenerate with under PBC. The
two states exhibit a repeated pattern of intertwining level-crossings
while their mean energy-difference to the ground state grows.  We do not
know of an intuitive way of understanding the spectrum for OBC from the
spectrum with PBC. Modifying the couplings at the boundary of the
chain by a multiplicative factor of $\lambda$ and varying $\lambda$
between 0 and 1 we have studied the cross-over from PBC to OBC. A
low-energy spectrum similar to the one for OBC is observed until
$\lambda \approx 0.9$.

\begin{figure}[ht]
  \centering
  \includegraphics[]{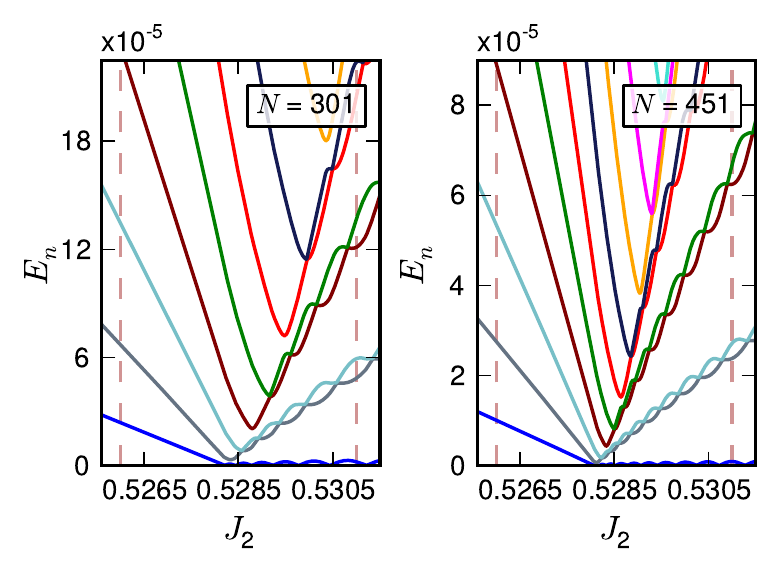}
  \caption{(Color online.) Spectrum with open boundary conditions for chains of 301
  and 451 sites. The values at which the scaling with $N$ is studied in
  Fig.~\ref{figscalingwithN} are indicated by dashed vertical lines.}
  \label{figspectrumtwoN}
\end{figure}

It is reasonable to ask if the Lifshitz point is a well defined point in
the spectrum. In order to answer this question we show results in
Fig.~\ref{figspectrumtwoN} for chains of length 301 and 451 sites for a
range of $J_2$ close to $J_2^L$.  As can be clearly seen, the
minima of the higher energy levels occur much closer to the first
level crossing of the ground state for $N=451$ than for $N=301$. In the
thermodynamic limit we expect the minima for all higher lying
levels to occur at $J_2^L$.

\begin{figure}[htb]
  \centering
  \includegraphics[]{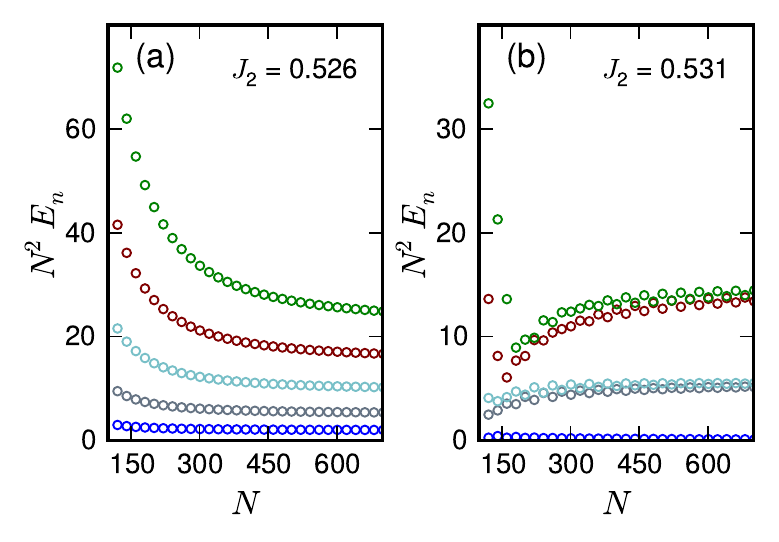}
  \caption{(Color online.) The energy scales proportionally to $N^{-2}$. 
  Plot of $N^2 \Delta E_n$ to the left of the transition (a)
  and to the right of the transition (b) for the first five
  excited states.}
  \label{figscalingwithN}
\end{figure}

We next focus on the scaling of the energy levels with $N$.  In
Fig.~\ref{figscalingwithN}(a) we show data for $N^2 E_n$ taken at $J_2 =
0.526$, to the left of the transition as indicated in
Fig.~\ref{figspectrum}~and~\ref{figspectrumtwoN}. 
As can be seen, it converges to a constant value indicating that for
this value of $J_2$ $E_n \propto N^{-2}$.  For the first excited state
this behavior is apparent for quite short chains already and it seems
plausible that for higher excited states longer chains would lead to the
same decay proportional to $N^{-2}$. This scaling is not surprising
since the soliton behaves like a massive particle in a box.  We
therefore expect the low energy spectrum to be approximated by $\hbar^2
k^2/(2\Delta_{sol})$ with $k=\pi n/2N$, $n=1,2..$, yielding the expected
scaling of the energies as $N^{-2}$.

In Fig.~\ref{figscalingwithN}(b) we show data taken on the other side of
$J_2^L$ at $J_2 = 0.531$ (again indicated in
Fig.~\ref{figspectrum}~and~\ref{figspectrumtwoN}).  For the smallest $N$
shown, the higher excited states still show signs of the transition at
this value of $J_2$. For short chains the second, third, forth and fifth
excited state thus have minimum in Fig.~\ref{figscalingwithN}(b). For
chains with more than roughly 160 sites we see of intertwining pairs of
states familiar from Figs.~\ref{figspectrum} and \ref{figspectrumtwoN}.
The average energy of the pair at big $N$ also scales proportionally to
$N^{-2}$.  We therefore conclude that sufficiently far away from the
transition point for the first few energy-states $\Delta E_n \propto
J_2/ N^2$.
\begin{figure}[ht]
  \includegraphics[]{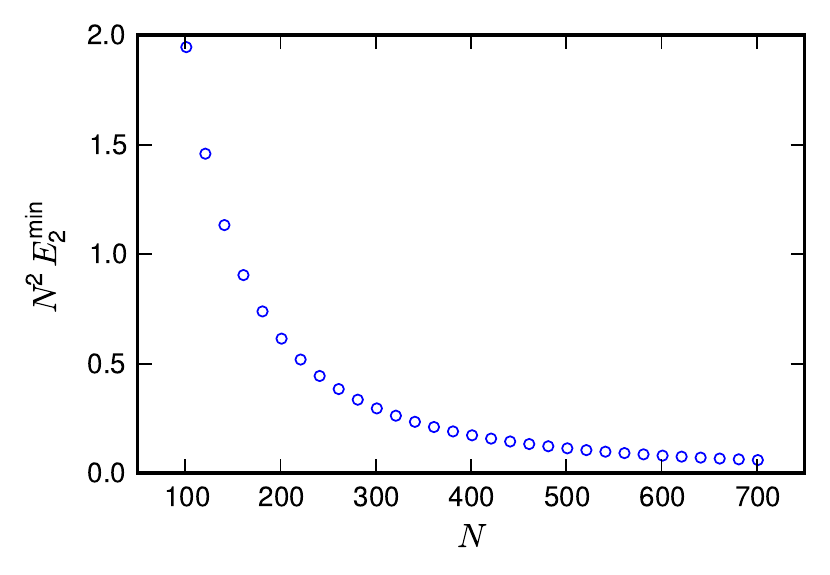}
  \caption{(Color online.) The energy at which the 2$^{\mathrm{nd}}$ excited state turns
  around goes to zero faster than $N^{-2}$. The value of $J_2$ of
  $E_2^\mathrm{min}$ was found up to $\Delta J_2 = 10^{-5}$.  The
  resulting uncertainty of the value of the minimum energy is smaller
  than the size of the symbols in the plot.}
  \label{figexcitedgap}
\end{figure}

In order to study the scaling of the spectrum at the Lifshitz point $J_2^L$
we focus on the minimum in the second excited state.
Although this minimum occurs at slightly different $J_2$ as $N$ is varied
it serves as the best possible definition of an excited energy scale at the Lifshitz point.
Specifically, we define the minimal energy-difference of the
ground state and the second excited state as $ E^{\mathrm{min}}_2 =
\mathrm{min}_{J_2}[E_2 - E_0]$.  Our results for $ E^{\mathrm{min}}_2$
are shown in Fig.\ref{figexcitedgap}.  As can be clearly seen in this
figure $ E^\mathrm{min}_2$ goes to zero faster than $N^{-2}$ violating
the simple scaling found elsewhere.
\begin{figure}[ht]
  \includegraphics[]{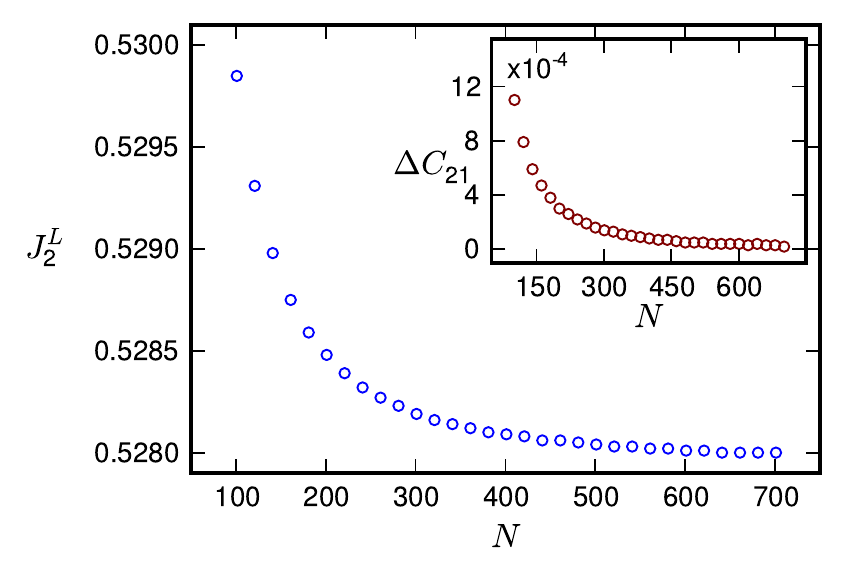}
  \caption{(Color online.)  Main panel: $J_2^L$ as a function of the
  length of the chain.  Inset: The difference between $J_2^L$ and the
  $C_2$ at which the minimum of the second excited state occurs.
  All values were determined up to $\Delta J_2 = 10^{-5}$. The resulting
  uncertainty of the value of the minimum energy is smaller than the
  size of the symbols in the plot and causes deviations from very smooth
  behavior.} 
  \label{figzerocrossing}
\end{figure}

We now turn to an estimate of the location of $J_2^L$ within the
variational approach.  The level-crossing of the first excited and the
ground state allows for an easy way to define the value of $J_2^L$ for a
given $N$.  As one could already see in the Figures~\ref{figspectrum}
and \ref{figspectrumtwoN}, $J_2^L$ varies slightly with the length of
the chain.  Our results are shown in Fig.~\ref{figzerocrossing} for
chains out to $N=701$.  The main panel in Fig.~\ref{figzerocrossing}
shows that $J_2^L$ converges to approximately $J_2^L=0.528$  as one
increases the length of the chain.  

The value of $J_2$ at which the second excited state has its minimum
also approaches $J_2^L$.  To show this we use the value of $J_2$ at
which the $n$-th state reaches its first minimum for a given $N$.  We
call this quantity $C_n$.  The inset in Fig.~\ref{figzerocrossing} shows
$\Delta C_{21} = C_2 - J_2^L$. As can be seen, this quantity approaches
0 and the minimum for big $N$ thus lies at the Lifshitz point.

\section{Incommensurate behavior in the entanglement entropy}
\label{sec:entang}
\begin{figure}[ht]
  \includegraphics[]{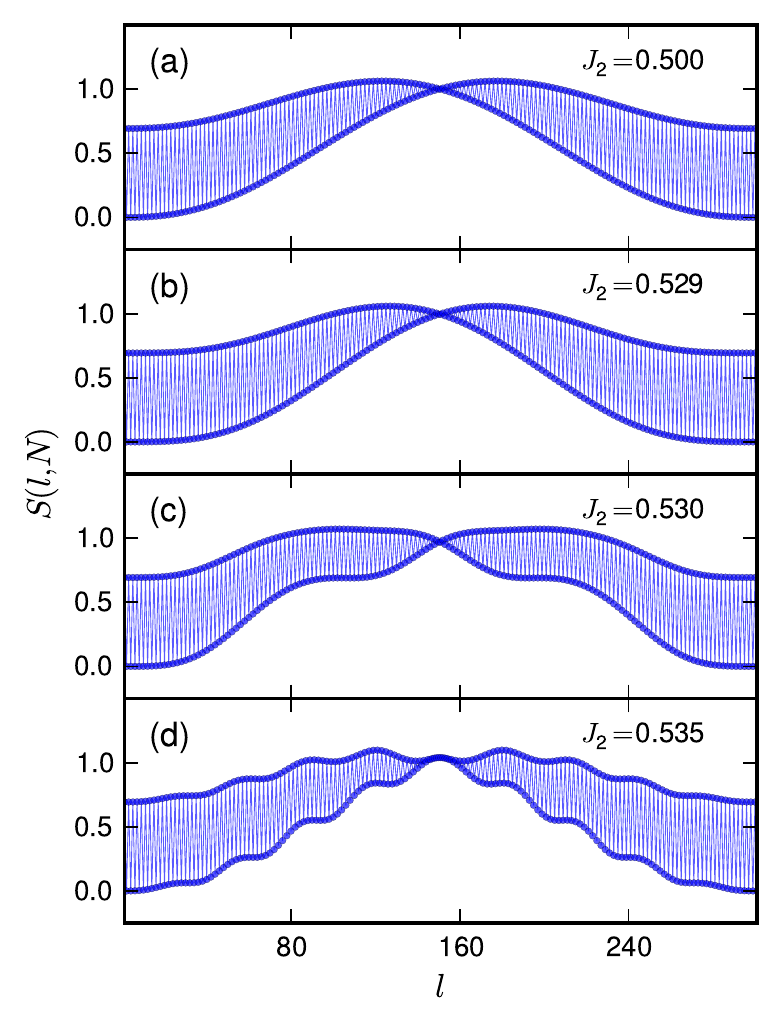}
  \caption{(Color online.) The entanglement entropy of a bipartition of
  an odd chain.  The data was taken for chain of 301 sites and  the
  variational subspace $R_0$ (see Sec.~\ref{sec:method}) was used. }
  \label{figentanglement}
\end{figure}

\begin{figure}[ht]
  \centering
  \subfloat[Even number of sites]{\includegraphics[]{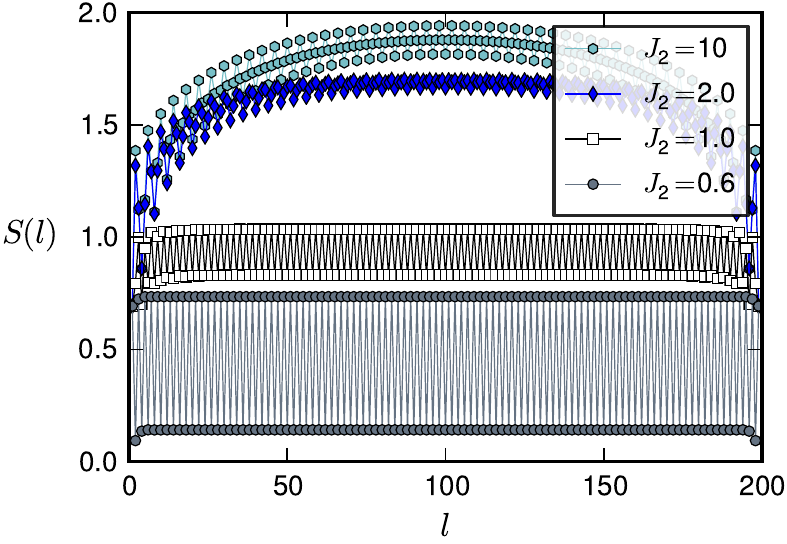}}\\
  \subfloat[Odd number of sites]{\includegraphics[]{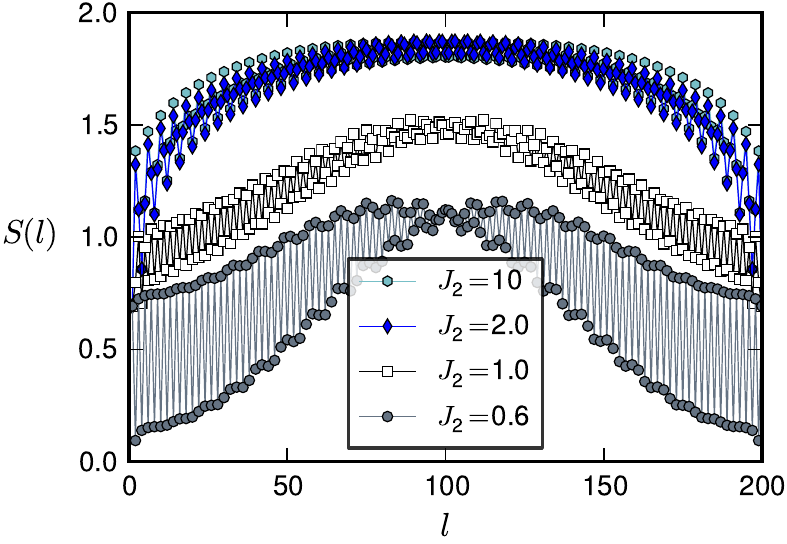}}

  \caption{(Color online.) The entanglement entropy of a bipartition of
  an even chain (200 sites) (a) and an odd chain (201 sites) (b). The
  data was obtained using DMRG with $m=256$ states kept.}
  \label{figdmrgentang}
\end{figure}

The scaling of the entanglement entropy at a (quantum) Lifshitz
transition has recently been the subject of
interest.~\cite{Fradkin_09,Rodney_2012} In free fermion models,
analogous to the spin chain model discussed here, the Lifshitz
transition is associated with a change in the topology of the Fermi
surface.  In one dimension new Fermi points appear at the Lifshitz
transition and, analogously, new patches appear in higher dimensional
models. If one associates a chiral conformal field theory with each
patch, it can be argued~\cite{Swingle_2010} that, when the number of
points (patches) increases by a factor $K$, the entanglement should be
multiplied with the same factor $K$.  For a free fermion model with next
nearest neighbor hopping, $t_2$, one expects the number of Fermi points to double at the
Lifshitz transition $t_2^L=1/2$ at half-filling with a corresponding
doubling in the entanglement entropy. This behavior is well confirmed in
numerical calculations.~\cite{Rodney_2012} 

In this section we discuss our results for the entanglement entropy
across the Lifshitz point in the odd length $J_1$-$J_2$ quantum spin
chain which is the quantum spin analogue of the model considered in
Ref.~\onlinecite{Rodney_2012}.  We study the entanglement in terms of
the von Neumann entanglement entropy of a sub-system $A$ of size $l$ and
reduced density-matrix $\rho_A$ defined by~\cite{Neumann27,Wehrl78},
\begin{equation}
  S(l,N)\equiv -\mathrm{Tr} [\rho_A\ln \rho_A]\ ,
  \label{eq:vNS}
\end{equation}
where $N$ again stands for the total system size. We consider
exclusively open boundary conditions.

If one uses the restricted space $R_0$,  which was introduced in
Sec.~\ref{sec:method}, as the variational subspace, one can also
calculate the entanglement entropy using the method employed in this
paper. \cite{soerensen_quantum_2007} Away from MG and Lifshitz points,
where the variational method is not reliable,  we complement  the
variational results with DMRG calculations.

We first discuss the variational results for odd length chains close to
the Lifshitz point shown in Fig.~\ref{figentanglement} for $N=301$.  The
entanglement entropy at the MG point for the odd length chain, shown in
Fig.~\ref{figentanglement}(a), has previously been discussed in detail.~
\cite{soerensen_quantum_2007} Since the entanglement entropy is very
directly connected to the wave function of the state, drastic changes of
the wave function should also be present in the entanglement entropy
when the Lifshitz point is reached. This is clearly the case as can be
seen in  Fig.~\ref{figentanglement}. As the Lifshitz point, $J_2^L$, is
reached, the entanglement entropy develops {\it plateaus}
(Fig.\ref{figentanglement}(c)).  As $J_2$ is increased more plateaus
appear (Fig.~\ref{figentanglement}(d)). For the free fermion model
studied in Ref.~\onlinecite{Rodney_2012} analogous oscillations in the
entanglement entropy are observed beyond $t_2>1/2$.  Because a different
subspace was used in the previous parts of this paper,  the transition
begins at  $J_2\sim0.529$  which is slightly higher than $J_2\sim 0.528$
which could be inferred from Fig.~\ref{figspectrumtwoN}.

For an even length system {\it no} such plateaus are visible (see
Fig.~\ref{figdmrgentang}(a)).  As $J_2\to \infty$ the entanglement {\it
increases} towards that of {\it two} independent gapless Heisenberg
chains as it must. A similar increase is seen for an odd number of sites
but with pronounced signatures of the incommensurability (see
Fig.~\ref{figdmrgentang}(b)). 

\section{The transition point}
\label{sec:trans}
The numerical value of $J_2^L$ for the Lifshitz point depends not only
on the length of the chain but also on the basis-set that one uses in
the variational calculation.  While this is a small  concern when one
looks at qualitative features, it is of course detrimental if one is
interested in a precise estimate of the Lifshitz point.  Just using the
different basis sets introduced in Sec.~\ref{sec:method} this is
evident. Using the smallest basis, $R_0$, for a chain with $N=301$
sites, we obtained $J_2^L\approx 0.5295$ (see
Fig.~\ref{figentanglement}) for the onset of oscillations in the
entanglement. This is a slightly bigger value than what was found in
Fig.~\ref{figspectrumtwoN}, $J_2^L\approx 0.528$, based on the calculations
with the larger basis $Z_s$.

DMRG can give us a more reliable estimate for $J_2^L$ at least for small
chains. For a chain with $201$ sites we found the first indications of
incommensurate behavior in the local magnetization at:
\begin{align}
  J_2^L \approx 0.538(1)
  \label{eqDMRGJ2}
\end{align}

We expect this estimate to depend on $N$ in roughly the same way as the
variational estimate does in Fig.~\ref{figzerocrossing}. If this is the
case, an eventual extrapolation to the $N\to\infty$ limit might change
this estimate by 0.0005 which is smaller than the uncertainty to which
we have determined the point.

\section{Conclusion}
We have studied incommensurability effects as they occur in the odd
length antiferromagnetic $J_1$-$J_2$ chain.  Even though no exact
ground state wave function is known at the MG point, $J_2=J_1/2$, this
point is the disorder point with minimal correlation length. The
Lifshitz point $J_2^L=0.538 J_1$ marks the onset of significant
modulations directly in the ground state $\langle S^z_i\rangle$ as well
as a shift in the ground state momentum. 
A series of inter-twining level crossings causing the shift in the
ground state momentum starts at the Lifshitz point.  The shift in the
ground state momentum and the associated modulations directly affect
the entanglement entropy which shows distinct plateaus developing for
$J_2>J_2^L$.

In realistic compounds with chain breaking impurities one would expect
half the chain segments to be of odd length. The experimentally well
studied compound CuGeO$_3$ has a $J_2\sim 0.36 J_1 < J_1/
2$.\cite{Riera_1995} If compounds with a $J_2$ in excess of $J_1/2$ can
be identified, it would be very interesting to experimentally look for
the odd length effects that we have detailed here. In particular, the
effects on the on-site magnetization shown in Fig.~\ref{figonsiteincm}
might be observable using NMR techniques or other local probes. 

\section*{ACKNOWLEDGMENTS}
We acknowledge many helpful discussions with Sung-Sik~Lee as well as with H.~Francis~Song, Marlon~Rodney and Karyn~Le~Hur.
This work is supported by NSERC.

\section*{References}

\end{document}